\documentclass[aps,prb,superscriptaddress,reprint,showpacs,floatfix]{revtex4-1}

\usepackage{tabularx}
\usepackage{float}
\usepackage{soul}
\usepackage{comment}
\usepackage{bm}
\usepackage{graphicx}
\usepackage{amsmath}
\usepackage{mathtools}
\usepackage{xcolor}
\usepackage{mathrsfs}
\usepackage[colorlinks=true,citecolor=blue]{hyperref}
\hypersetup{colorlinks=true,citecolor=blue,linkcolor=blue,urlcolor=blue}

\usepackage{chngcntr}

\begin{document}

\title{Microscopic theory of exciton and trion polaritons in doped monolayers of transition metal dichalcogenides}

\author{Y. V. Zhumagulov}
\email{yaroslav.zhumagulov@gmail.com}
\affiliation{University of Regensburg, Regensburg 93040, Germany}
\affiliation{Department of Physics and Engineering, ITMO University, St. Petersburg, 197101, Russia}

\author{S. Chiavazzo}
\affiliation{Department of Physics and Astronomy, University of Exeter, Stocker Road, Exeter EX4 4QL, United Kingdom}

\author{D. R. Gulevich}
\affiliation{Department of Physics and Engineering, ITMO University, St. Petersburg, 197101, Russia}

\author{V. Perebeinos}
\affiliation{Department of Electrical Engineering, University at Buffalo, The State University of New York, Buffalo, New York 14260, USA}

\author{I. A. Shelykh}
\affiliation{Department of Physics and Engineering, ITMO University, St. Petersburg, 197101, Russia}
\affiliation{Science Institute, University of Iceland, Dunhagi-3, IS-107 Reykjavik, Iceland}

\author{O. Kyriienko}
\affiliation{Department of Physics and Astronomy, University of Exeter, Stocker Road, Exeter EX4 4QL, United Kingdom}

\begin{abstract}
We study a doped transition metal dichalcogenide (TMDC) monolayer in an optical microcavity. Using the microscopic theory, we simulate spectra of quasiparticles emerging due to the interaction of material excitations and a high-finesse optical mode, providing a comprehensive analysis of optical spectra as a function of Fermi energy and predicting several modes in the strong light-matter coupling regime. In addition to the exciton-polaritons and trion-polaritons, we report additional polaritonic modes that become bright due to the interaction of excitons with free carriers. At large doping, we reveal strongly coupled modes reminiscent of higher-order trion modes that hybridize with a cavity mode. We also demonstrate that rising the carrier concentration enables to change the nature of the system's ground state from the dark to the bright one. Our results offer a unified description of polaritonic modes in a wide range of free electron densities. 
\end{abstract}

\maketitle


Monolayers of transition metal dichalcogenides (TMDC) represent a class of two-dimensional (2D) materials with remarkable optical properties. They are direct bandgap semiconductors with hexagonal Brillouin zone, with two nonequivalent valleys at the K and K$^\prime$ points~\cite{Mak2010,Kormanyos2015,Miwa2015}.
Due to large spin-orbit interaction, they experience spin-valley locking, which opens the way for spin and valleytronic applications~\cite{Mak2012}. Moreover, the direct bandgap type and relatively large effective masses of electrons and holes lead to the formation of robust bright excitons~\cite{Splendiani2010}. Due to large exciton binding energy, they dominate an optical response of TMDC materials even at room temperature \cite{Steinleitner2017,Wang2018}. Also, thanks to peculiar Coulomb interaction screening in 2D~\cite{Keldysh1979,Glazov2018,Huser2013} TMDC excitons have a non-Rydberg energy spectrum~\cite{Chernikov2014}. Combined with the remarkable compatibility of these materials with various semiconductor/dielectric platforms, this makes them promising candidates for the development of various nanophotonic components \cite{Xia2014}, including logic circuits \cite{Radisavljevic2011,Wang2012}, phototransistors \cite{LopezSanchez2013}, and light sensors \cite{Perkins2013} as well as light-producing and harvesting devices
\cite{Feng2012,Pospischil2014}. Studies of defects in TMDC monolayers and increased confinement also led to efficient single-photon emitters \cite{Flatten2018,Baek2020}.

One of the perspective fields of the application of TMDC monolayers is polaritonics. Exciton polaritons are hybrid light-matter quasiparticles formed due to the strong coupling between excitons and a tightly-confined optical mode (reviewed in Ref.\,[\onlinecite{Kockum2019}]). Possible realizations include coupling to photonic crystal cavities, surface plasmons, nanoparticle resonances, open fiber cavities, and planar microcavities based on distributed Bragg reflectors (see the full polariton panorama in Ref.\,[\onlinecite{Basov2020}]).
In this configuration, a plethora of quantum collective phenomena is observed at surprisingly high temperatures, which include polariton lasing \cite{Kasprzak2006,Schneider2013,Ballarini2017} and emergent polariton fluid behavior \cite{Carusotto2013}, and nontrivial polariton lattice dynamics \cite{Askitopoulos2013,Ohadi2017,Gao2018,Mitki2018,Kyriienko2019}. These phenomena pave the way for ultrafast polariton-based nonlinear optical integrated devices \cite{Liew2011,Ohadi2015,Kyriienko2016,Askitopoulos2018,Opala2019}.

The strong light-matter coupling regime is achieved when the exciton-photon coupling, characterized by the vacuum Rabi splitting $\Omega$, overcomes losses, $\Omega > \kappa, \gamma$~\cite{Kockum2019}. The latter stem from the finite transmittance of the photonic cavity mirrors ($\kappa$) and a finite non-radiative lifetime of the excitons ($\gamma$). TMDC monolayers are very promising in this context. Indeed, as compared to conventional semiconductor materials, TMDC excitons have substantially higher binding energies. This makes the resulting TMDC polariton stable even at room temperatures and provides high optical oscillator strengths \cite{Schwarz2014,Liu2014,Dufferwiel2015,Lundt2016,Schneider2018,Zhang2018,Sthrenberg2018,Yankovich2019,Gonalves2020}.
\begin{figure}[t]
    \centering
    \includegraphics[width=0.95\linewidth]{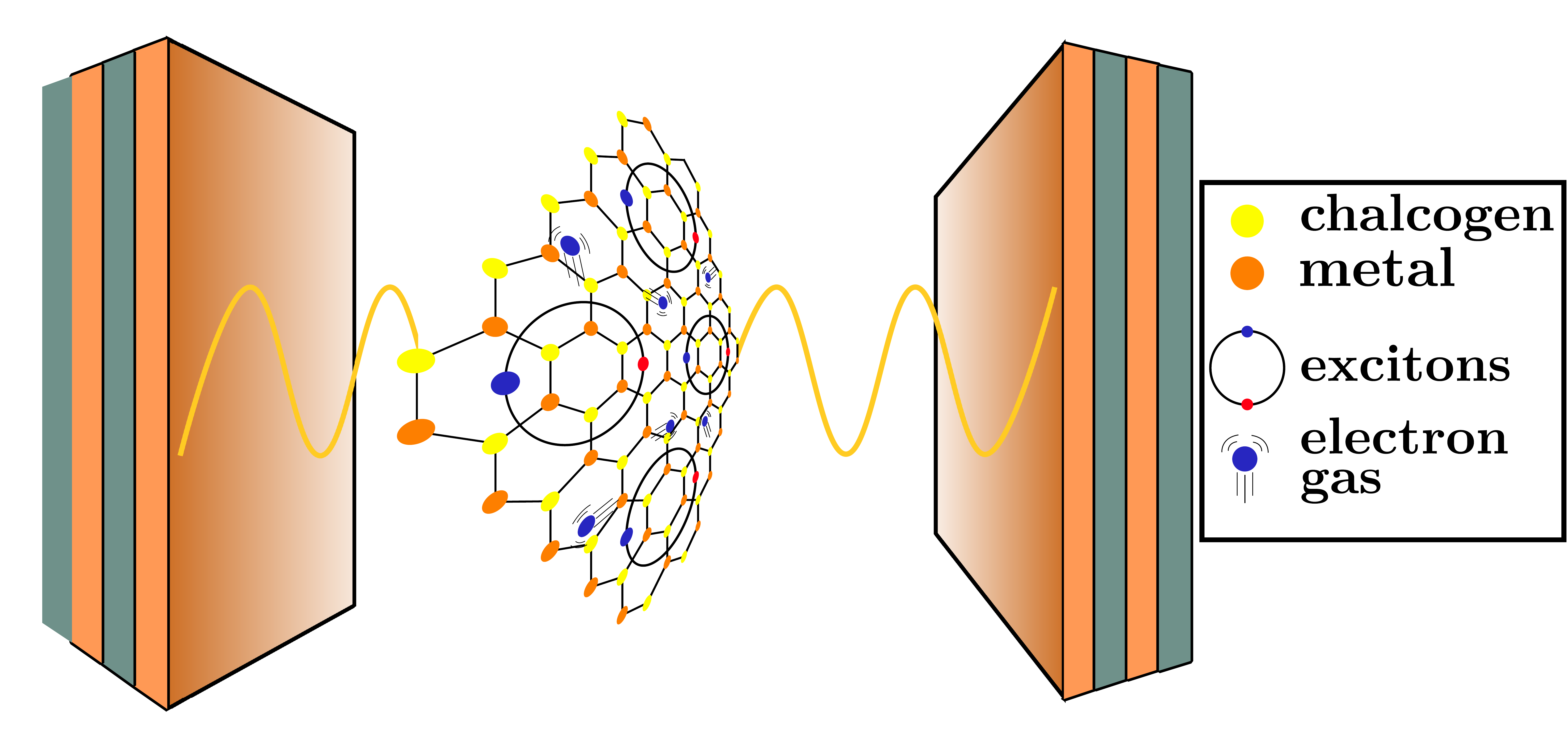}
    \caption{\textbf{Sketch of the system.} Transition metal dichalcogenide monolayer in an open microcavity with adjustable cavity frequency. The hexagonal TMDC lattice hosts bound excitonic states (circles) and a gas of electrons (blue bearded dots). Yellow and orange dots are chalcogen and metal atoms, respectively.
    }
    \label{fig:sketch}
\end{figure}

The interaction of excitons with free electron gas, which can be present in a TMDC monolayer due to external doping, substantially modifies its optical response. The position of the excitonic line is shifted due to the effects of the bandgap renormalization \cite{Ugeda2014,Chernikov2015a,Chernikov2015,Withers2015,Raja2017}. In addition, complimentary screening and effects of Pauli blocking change the structure of the excitons, making them less compact and reducing the corresponding oscillator strength and Rabi splitting \cite{Shahnazaryan2020}. Other absorption peaks appear due to the brightening of an intervalley exciton \cite{Zhumagulov2020}. 
The interaction between excitons and an electron gas leads to the appearance of charged quasiparticles being trions  \cite{Mak2012,Ross2013,Singh2016,Courtade2017,Lundt2018,Rana2020} or exciton polarons~\cite{Sidler2016,Efimkin2017,Ravets2018,Chang2018,Tan2020}. The presence of such quasiparticles can qualitatively modify the light-matter coupling and completely reshape the polaritonic spectrum \cite{Emmanuele2020,Kyriienko2020}. Moreover, various nonlinear effects~\cite{Shahnazaryan2017,Barachati2018,Kyriienko2020,Kravtsov2020,Tan2020,Stepanov2021,Li2021a,Li2021b,BastarracheaMagnani2021,Denning2021a,Denning2021b} arise from the phase-space filling, hybridization with light exciton-exciton and exciton-electron scattering, which offers a route towards polariton condensation~\cite{AntonSolanas2021} and the emergence of spontaneous coherence in monolayers strongly coupled to light \cite{Shan2021}.

Here, we develop the quantitative microscopic theory for exciton-polariton states in TMDC monolayers with different doping levels. We consider a TMDC monolayer placed in an optical microcavity  (Fig.~\ref{fig:sketch}). We solve the three-body problem by exact diagonalization of the Hamiltonian in a three-particle basis set~\cite{Deilmann2016,Drppel2017,Torche2019,Tempelaar2019,Zhumagulov2020,Zhumagulov2020a}, where an eigenmode solution is available at small and high free electron concentrations. We observe that an additional dressed-exciton mode appears at small doping, apart from the dominant exciton and trion polariton contributions. Qualitatively, its formation can be explained as a result of the brightening of an intervalley exciton in the presence of a Fermi sea, where free electrons contribute an additional momentum necessary for the optical recombination of the intervalley exciton hole with the doping electron. This couples strongly to the cavity mode and leads to a dipolariton-like spectrum. We observe a strongly modified spectrum at large doping at high frequencies, attributing it to the exotic excited-trion state. Our work lays the solid background for understanding the spectrum of TMDC monolayer excitations at the strong light-matter coupling.


\section*{The model}

\subsection{Single-particle states}

We start by describing the microscopic theory for polaritons in the electron-doped transition metal dichalcogenide monolayers. We consider a generic description valid for different TMDC materials. In our calculations, we use MoS$_2$ as an example. First, we calculate the single-particle states of a single TMDC monolayer. For this, we use the massive Dirac model~\cite{Xiao2012} that is well suited to describe the single-particle band structure in the vicinity of $K$ and $K^\prime = -K$ points of the Brillouin zone. The massive Dirac Hamiltonian reads as ($\hbar = 1$)
\begin{align}
    \hat{\mathcal{H}}_0&=  v_{\mathrm{F}} \hat{s}_0\otimes \big( \tau k_x\hat{\sigma}_x+k_y\hat{\sigma}_y + \frac{\Delta}{2} \hat{\sigma}_z \big) \notag \\ 
    &+ \tau \hat{s}_{z} \otimes \big( \lambda_c \hat{\sigma}_{+} + \lambda_v \hat{\sigma}_{-} \big),
\label{HDirac}
\end{align}
where $\Delta$ is the bandgap, $v_F$ is the Fermi velocity, $k_{x,y}$ are the electron crystal momentum component, $\tau$ takes into account for the valley degrees of freedom and $\lambda_{c,v}$ are the SOC (spin-orbit coupling) constants. In addition, $\hat \sigma_{x,y,z}$ are pseudospin Pauli matrices acting in the band subspace, $\hat \sigma_{\pm} = (\hat \sigma_x \pm i \hat{\sigma}_y)/2$ are raising and lowering operators, while $\hat s_{0},\hat s_{z}$  are identity and spin Pauli Z matrix in the electron spin subspace, respectively. The role played by the SOC is included in the last term of Eq.~\eqref{HDirac}.

We note that the Hamiltonian~\eqref{HDirac} matches the density functional theory calculations at low energies~\cite{Zollner2019,Kormanyos2015}. Specifically,
trions are formed by the states near the $K$ and $K^{\prime}$ points and the contribution from states away from the corners of the Brillouin zone are negligible~\cite{Zhumagulov2020}.
Despite the relative simplicity, the model captures all relevant phenomena, including the interplay between the spin and the valley degrees of freedom. 
\begin{figure}
    \centering
    \includegraphics[width=0.95\linewidth]{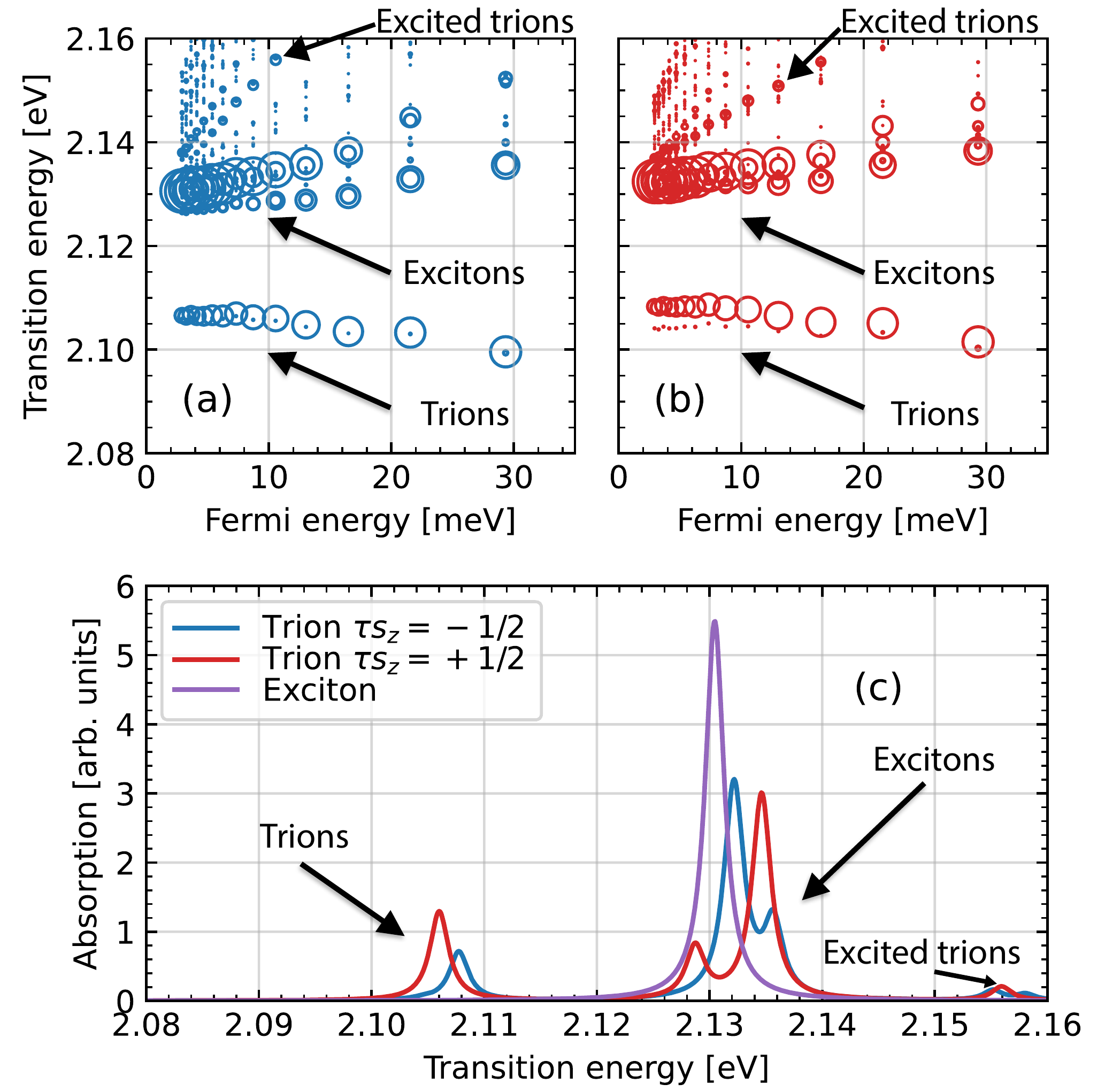}
    \caption{\textbf{Trionic and excitonic spectra}. (a) Doping dependence of the transition energy for three-body states shown by blue circle-shaped markers spin-valley index $\tau s_z=-1/2$ and (b) by $\tau s_z=+1/2$. The size of each marker is proportional to its optical oscillator strength. 
    (c) The absorption spectrum of an MoS$_2$ monolayer. The purple curve shows the absorption spectrum in the limit of zero doping level, where a single peak corresponds to 1s exciton. The blue curve corresponds to the absorption spectrum of three-particle states with $\tau s_z=-1/2$ at the level of doping equal to $E_F=10.56$~meV. The red curve shows the same dependence for $\tau s_z=+1/2$  three-particle states.}
    \label{fig:levels}
\end{figure}

To account for the effects of the dielectric environment, we introduce the bandgap dependence on the substrate and monolayer properties. The bandgap renormalization enters as the scissor shift $\delta$ for the bandgap $\Delta = \Delta_0 + \delta$ that modifies the bare monolayer bandgap $\Delta_0$. 
The Fermi velocity is then adjusted as $v_{F}= \sqrt{\Delta/2m}$ to preserve the effective mass amplitude~\cite{Waldecker2019}. For the case of MoS$_2$ we consider the effective masses of electrons and holes to be $0.52 m_0$ ($m_0$ is the free electron mass), bandgap $\Delta_0$ of $2.087$~eV, $\lambda_c$ of $1.41$~meV, $\lambda_v$ of $74.6$~meV, the bulk dielectric constant $\varepsilon_0$ of 14, the dielectric thickness of $6$~\AA, and scissor operator $\delta$ equal to 0.64~eV.


\subsection{Many-particle states}

\begin{figure}
\includegraphics[width=0.95 \linewidth]{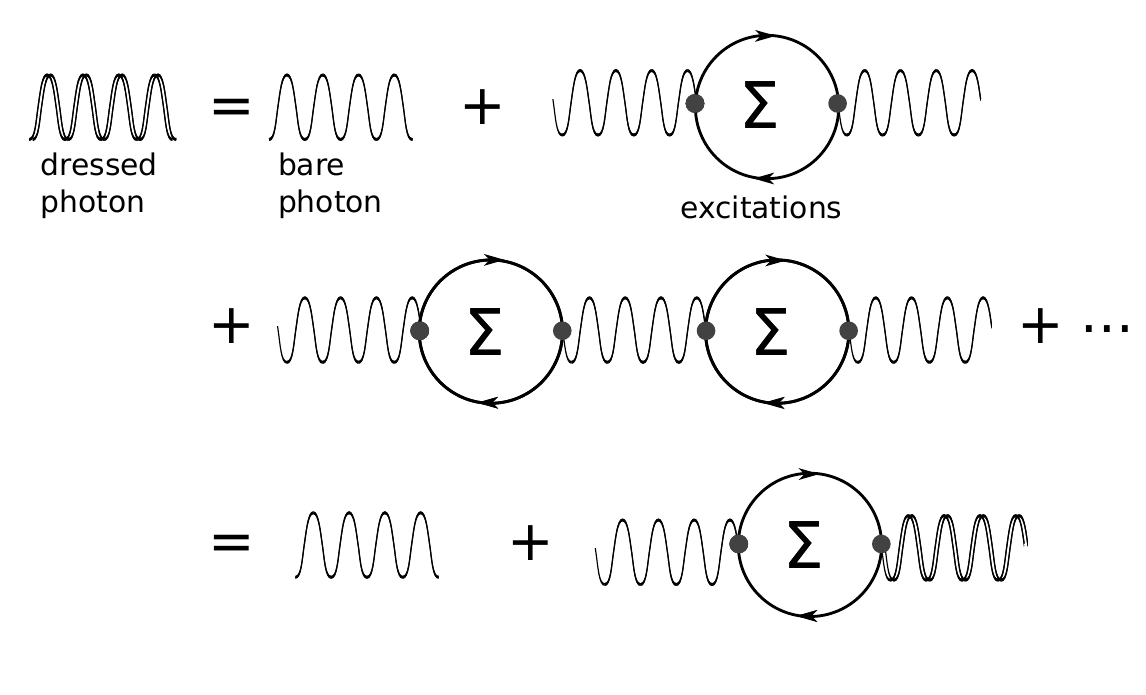} 
\caption{\textbf{Diagrammatic Dyson equation.} The bare cavity photon propagator (thin wavy curve) is dressed via multiple absorption and re-emission processes by TMDC excitations ($\Sigma$). The dressed cavity photon Green's function is depicted by a thick wavy curve. The solution of this diagrammatic equation is written in Eq.~\eqref{eq:Dyson} in the algebraic form.
}
\label{fig:Dyson}
\end{figure}

We introduce the interaction between electrons and holes and diagonalize the many-body three-particle Hamiltonian. In TMDC monolayers, the intravalley interaction is described by the screened potential in the Rytova-Keldysh form.~\cite{Rytova1967,Keldysh1979,Cudazzo2011} The Fourier transform of the screened Coulomb between particles in the same valley reads $W(q)=V(q)/[\varepsilon_{+}(1 + r_0 q)]$, where $V(q)=2\pi e^2/q$ is the bare Coulomb potential in 2D, and $q$ is the magnitude of exchanged wavevector. We use the average substrate permittivity $\varepsilon_{+} = (\varepsilon_{2}+\varepsilon_{1})/2$ and the screening length $r_0=\varepsilon_{0} d/2$.~\cite{Berkelbach2013,Cho2018} The screened Coulomb potential for the intervalley processes reads $W(q)=V(q)/\varepsilon_{0}$.\cite{Florian2018}  We neglect the wetting layer, vacuum spacing between the monolayer and substrate, which otherwise  will modify the potential.~\cite{Florian2018,Shahnazaryan2019}

\textit{Excitons and trions in the presence of doping.} 
To calculate the spectrum of three-body quasiparticles in the monolayer and find the corresponding eigenstates, we define operators $\hat{T}_{\nu}^\dagger$ ($\hat{T}_{\nu}$) that create (annihilate) states $\nu$ with two electrons in the conduction band (labeled as $c_{1,2}$) and a hole in the valence band (labeled as $v$). The corresponding basis is spanned by linear superpositions
\begin{align}
\label{eq:T_basis}
    \hat{T}_{\nu}^\dagger \left| {\O} \right\rangle  = \sum_{c_1,c_2,v} A_{\nu, c_1 c_2 v} a_{c_1}^\dagger  a_{c_2}^\dagger a_{v}^\dagger \left| {\O} \right\rangle ,
\end{align}
where $\hat{a}_{b}^\dagger$ is the fermion creation operator for a state in band $b$, and $A_{\nu, c_1 c_2 v}$ denote the amplitudes. Note that $c_{1,2}$ are referring to the conduction band, and we use the usual anticommutation relations $\{ \hat a^\dagger_{c_1}, \hat a^\dagger_{c_2}\} = \delta_{c_1,c_2}$. The valence band is completely separated, and we assume the relation $\{\hat a_{c_{1,2}}, \hat a_v\} = 0$ to hold. At this point we stress that the introduced three-body operator can describe both a tightly bound state of two electrons and holes, as well as an electron-hole pair (exciton) in the presence of another electron.
\begin{figure*}
\includegraphics[width=0.95 \linewidth]{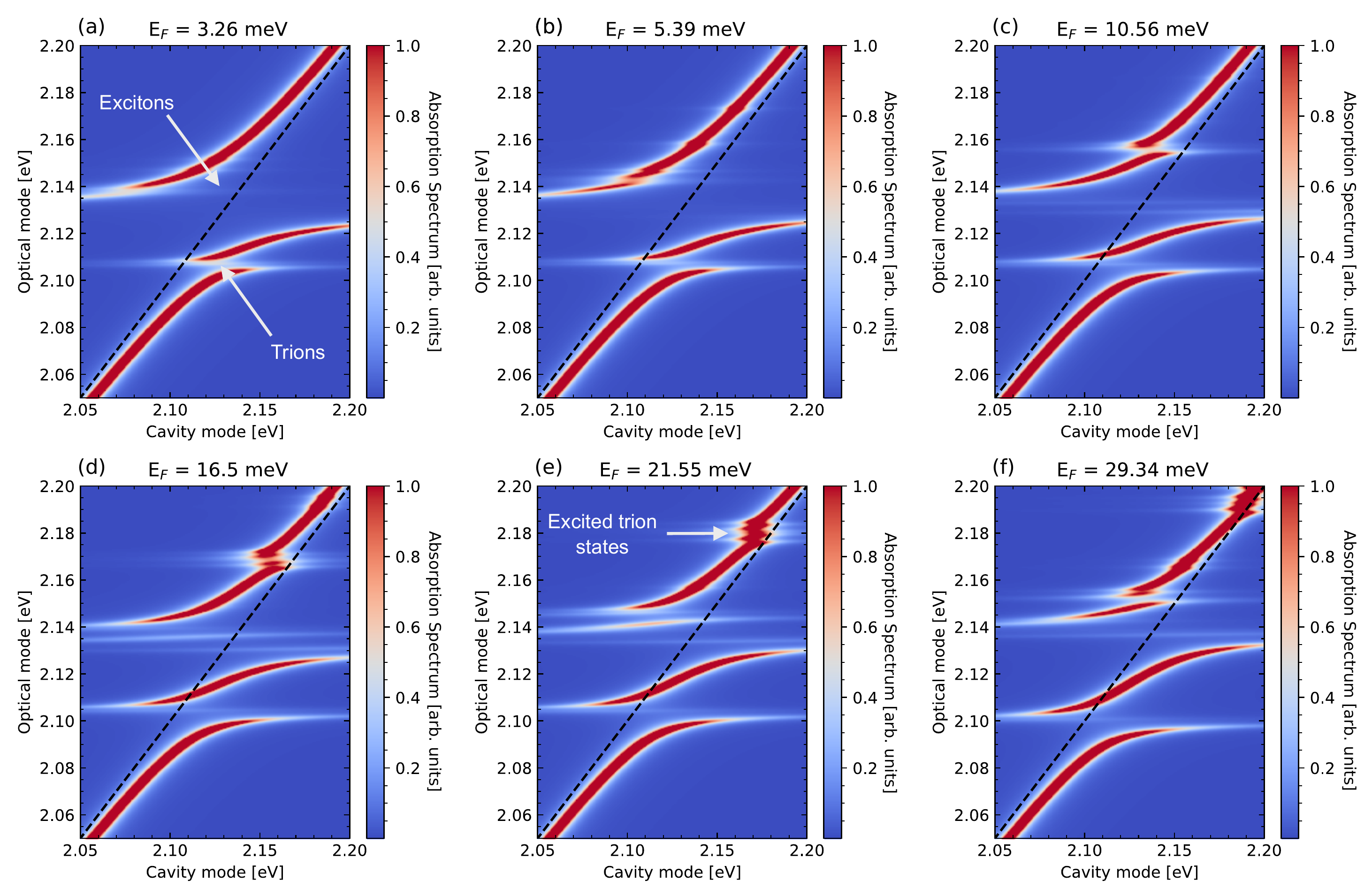} 
\caption{\textbf{Polaritonic spectra}. Absorption is shown as a function of detuning for increasing Fermi energy (left-to-right). We consider $\sigma_+$ polarization for the cavity mode. Color bars define the intensity in arbitrary units. In all the spectra, from low ($3.26$~meV) to high ($29.34$~meV) Fermi energy, we see the emergence of trion modes at $\sim 2.1$~eV with considerable Rabi splitting ($\sim 10$~meV), being significantly larger than the broadening $\gamma = 1$~meV}. The exciton-polariton features the appearance of the additional mode inside the polaritonic gap. At high free electron concentration, the additional splitting is attributed to the higher-order trion state.
\label{fig:spectra}
\end{figure*}

Using the generalization of the Tamm-Dancoff approach \cite{Deilmann2016,Drppel2017,Torche2019,Tempelaar2019,Zhumagulov2020,Zhumagulov2020a}, we decompose the full many-body Hamiltonian $\hat H$ in the three-particle state basis and solve the corresponding eigenvalue problem (see Methods for the details).
We obtain the set of low-energy eigenstates and corresponding energies to high precision, which allows describing the optical properties of the system. To classify the states in trionic basis, we look at the product $\tau s_z$, with $\tau = \tau_{c_1} + \tau_{c_2} - \tau_{v}$ being the total valley index and $s_z = s_z^{c_1} + s_z^{c_2} - s_z^{v}$ being the total spin index. The total spin $s_{z}$, in general, can have magnitudes $1/2$ and $3/2$. We restrict our analysis to $s_z = \pm 1/2$ spin projections as states that participate in optical recombination processes and neglect spin-$3/2$ states. The case $\vert s_z \tau \vert = 1/2$ is a necessary condition for the trion state to be bright. In the numerical procedure, we also account for the symmetries stemming from the conservation of total spin $s_z$ and total momentum ${\bf k} = {\bf k}_{c_1} + {\bf k}_{c_2} - {\bf k}_v$.

We model the presence of doping by introducing Pauli blocking through $k$-space discretization. The $k$-space grid is arranged as a mesh of $N\times N$ points, with the area of the primitive cell being $\Omega_0$. The doping density can be introduced as $n=g_v g_s/(\Omega_0 N^2)$, where $g_s$ and $g_v$ are the spin and valley degeneracies, respectively. The corresponding Fermi energy is equal to $E_{\mathrm{F}}= \delta k^2/2m$, where $\delta k = 4\pi/(\sqrt{3} a N)$ is the interval between k-points, and $a$ is a lattice constant for the hexagonal lattice. The described approach allows obtaining both trion and exciton states in the presence of doping from the  three-particle Hamiltonian, and in the zero doping limit  ($N\to\infty$) recovers the results from the Bethe-Salpeter equation for the excitonic states~\cite{Zhumagulov2020}. Therefore, we treat trionic and dressed excitonic states on equal footing and study polaritonic modes in the broad range of doping levels.

Together with eigenenergies of qus siparticles in the TMDC monolayer, we obtain the eigenstates and access various system observables.  Specifically, we calculate dipole matrix elements $\mathbf{D}_{\nu}$ for interband transitions for the three-body state $\nu$. Dipole moments are labeled by the corresponding band and read
\begin{equation}
    \mathbf{D}_{\nu} = \sum_{c_1,c_2,v} A_{\nu, c_1 c_2 v} (\textbf{d}_{vc_1}\delta_{c,c_2} - \textbf{d}_{vc_2}\delta_{c,c_1}), 
    \label{D}
\end{equation}
where $\mathbf{d}_{vc}$ is the single-particle dipole matrix element for bands $v$ and $c$, and $\delta_{c,c_1}$ is a Kronecker delta. It should be noted that from the above expression, it follows that trions with a finite moment also have coupling with optical cavity mode. However, in our work, based on the assumption of zero temperature, we have calculated trions only with a moment equivalent to the moment of the centers of the $K$ and $K^\prime = -K$ valleys \cite{Zhumagulov2020a}.

We present the results for a TMDC monolayer (MoS$_2$) in the case of far-detuned optical mode (Fig.~\ref{fig:levels}). The three-particle states are obtained by solving Eq.~\eqref{eq:matrix-equation} at different levels of doping, i.e. Fermi energy $E_\mathrm{F}$. In Fig.~\ref{fig:levels}(a), we plot the first 64 eigenenergies as a function of $E_\mathrm{F}$. We observe two dominant sets of modes at low energies for both spin-valley indices $\tau s_z = +1/2$ and $\tau s_z = -1/2$. The lowest set is attributed to trion-bound states. Trion energies, at around $2.15$~eV at small free electron concentrations ($E_\mathrm{F} = 2.93$~meV), decrease sublinearly with increasing $E_\mathrm{F}$ to $2.1$~eV at large concentrations. With the increasing doping level, the redshift of the trion line can be attributed to the increase of the excitonic polarizability, which is necessary for the binding of an additional electron.  The size of the markers denotes the magnitude of dipole moments and shows the gradual growth of the trion oscillator strength. As expected, the excitonic modes, which are located $\sim 25$~meV above trions, reveal opposite behavior with respect to trions, as they have the largest optical dipole moment at low concentrations. At the same time, with increasing $E_\mathrm{F}$ we observe gradual redistribution of the oscillator strength between different excitonic modes as well as their blueshifts [Fig.~\ref{fig:levels}(a)], which can be attributed to the effects of screening and Pauli blocking \cite{Efimkin2017,Shahnazaryan2020}. We also note the appearance of a higher-lying band of states that gain significant optical dipole moments.

To plot the photoluminescence (PL) in the weak coupling regime, we assign the final lifetime for each quasiparticle and plot the PL spectrum as a set of Lorentzians [Fig.~\ref{fig:levels}(b)]. Choosing a large doping limit, we observe the strong trion mode, the shifted exciton peak, and an additional brightened quasiparticle at 2.155 eV \cite{Zhumagulov2020a}.


\subsection{Light-matter coupling}

Next, we consider a planar photonic microcavity system with a doped monolayer of TMDC placed in the antinode (Fig.~\ref{fig:sketch}). Confined cavity photons tuned in resonance with TMDC modes will couple strongly to excitonic and trionic optical transitions, resulting in a polaritonic spectrum with substantial modifications. 
Recently, the question of non-perturbative ab initio simulations of exciton–polariton in TMDCs was addressed in Ref.\,[\onlinecite{Latini2019}]. To describe TMDC polaritons at increasing doping, we use the diagrammatic approach. Starting from the bare Green's function of cavity photons $G_0(\omega,\textbf{q})$ with wavevector $\textbf{q}$, we obtain polaritonic from the photonic Green's function dressed by excitation in the system. This process is shown diagrammatically in Fig.~\ref{fig:Dyson}, where multiple absorption and re-emission of a cavity photon lead to the mode hybridization~\cite{Levinsen2019,Kyriienko2012}. Notably, this is a non-perturbative approach where the dressed propagator can be found exactly by solving the corresponding Dyson equation. In the algebraic form the solution reads
\begin{equation}
\label{eq:Dyson}
G(\omega,\textbf{q})=\frac{G_0(\omega,\textbf{q})}{1-\Sigma(\omega,\textbf{q})G_0 (\omega,\bf{q})},  
\end{equation}
where
\begin{equation}
G_0(\omega,\textbf{q})= \frac{2\omega_\mathrm{cav}(\textbf{q})}{\omega^2-\omega_\mathrm{cav}^2(\textbf{q})+2i\gamma_\mathrm{cav} \omega_\mathrm{cav}(\textbf{q})},
\end{equation}
and $\omega_\mathrm{cav}(\textbf{q})=\sqrt{\omega_\mathrm{cav}^2 + c^2\textbf{q}^2}$ with $\omega_\mathrm{cav}$ being the resonant frequency of a cavity at normal incidence, $\gamma_\mathrm{cav}$ the broadening of the photonic linewidth due to a finite transmission coefficient of the cavity mirrors. The self-energy $\Sigma(\omega,\textbf{q})$ describes material excitations and accounts for the interaction vertices defined by the strength of light-matter interaction. It reads
\begin{equation}
\Sigma(\omega,\textbf{q})=\sum_{\nu=1}^d\frac{|\Omega_{\nu, \mathbf{q}}|^2}{\omega-\epsilon_\nu(\textbf{q})+i\gamma_\nu},
  \end{equation}
where $\Omega_\nu$ are Rabi frequencies of transitions, $\epsilon_\nu(\textbf{q})$ are their resonant frequencies, and $\gamma_\nu$ are non-radiative broadenings, depending on the quality of a sample. In total, we account for $d=64$ transitions coming from the exact diagonalization and presented in Fig.~\ref{fig:levels}(a). The broadenings are chosen based on typical experimentally achievable parameters.

The light-matter interaction is defined by the optical dipole moments of transitions and the amplitude of the electric field (or vector potential) for the cavity mode. The associated Rabi frequencies are
\begin{align}
\label{eq:Rabi_def}
    \Omega_{\nu, \mathbf{q}} = -2i \mathbf{D}_{\nu} \sqrt{\frac{\omega_\mathrm{cav}(\mathbf{q})}{2 \varepsilon \varepsilon_0 A}},
\end{align}
where $A$ is the cavity area, and $\mathbf{D}_{\nu}$ are dipole matrix elements of the transitions given by Eq.~\eqref{D}. We note that the confinement of quasiparticles, as well as their physical origin, are already accounted for in $\mathbf{D}_{\nu}$. Thus, this leads to the final result does not depend on the area of the sample $A$, but instead depends on the carrier concentration (see discussion in Ref.\,[\onlinecite{Combescot2003}]). For the dominant coupling to the exciton mode, the Rabi frequency is enhanced due to reduced exciton volume (area), independent of the number of particles.  For the case of trion-dominant modes, the Rabi frequency is enhanced at increasing electron concentration (being area independent), which will analyze directly from the microscopic calculations.

An important consequence of cavity photon dressing by TMDC excitations corresponds to hybridization of these modes, as can be seen from the modification of poles of $G(\omega,\mathbf{q})$ for couplings $|\Omega_{\nu,\mathbf{q}}|$ that dominate broadenings $\gamma_{\nu}$. The imaginary part of $G(\omega, \mathbf{q})$ then represents the absorption spectrum of the polaritonic system (up to scaling constant).\cite{Kyriienko2012} In the next section, we proceed to calculate the polaritonic spectrum and discuss the results.
\begin{figure}
    \includegraphics[width = 0.95\linewidth]{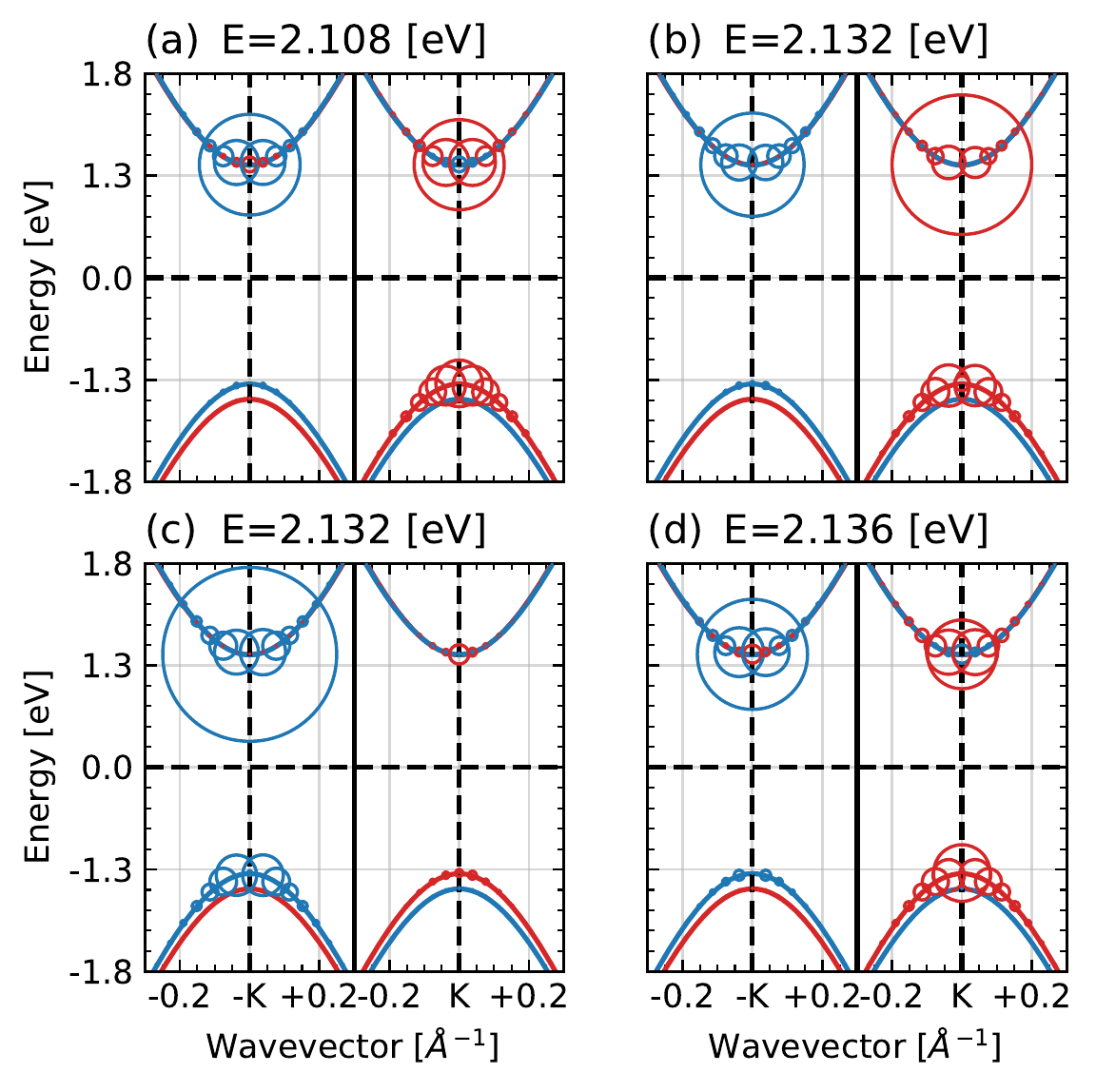}
    \caption{\textbf{Trion-polariton properties.} (a) Trion polariton energy splitting as a function of the square root of electron density (red circles and blue curve). We confirm the linear increase of splitting at low densities $E_\mathrm{F} \leq 16.5$~meV (cf. orange line). However, at large free electron densities, we see the growth becomes sublinear in $\sqrt{n_e}$. (b) Intensity profile for polaritonic spectrum at large doping ($E_\mathrm{F} = 29.34$~meV). The cavity mode energy is tuned to $2.11$~eV, being in resonance with the trion mode. The first two consecutive peaks correspond to trion-polaritons, with the Rabi splitting of $18$~meV. 
    }
    \label{fig:contributions}
\end{figure}


\section*{Results and discussion}

We consider a cavity polariton setup (see the sketch in Fig.~\ref{fig:sketch}) where excitons and trions couple to the cavity mode with a tunable frequency. The system can be conveniently realized in open cavities, where 0D cavity resonance can be shifted by cavity length change with a piezodrive, which allows coupling to excitations at energies $\sim 2$~eV and resulting anticrossing of various modes. Using the developed theory and extensive numerical calculations, we present polaritonic spectra for varying TMDC doping. We concentrate on the gated MoS$_2$ monolayer case and cover a broad range of densities. It is convenient to present the doping level using the Fermi energy $E_\mathrm{F}$ for the free electron gas. We start with an intrinsic value of $E_{\mathrm{F}} = 2.92$~meV and move up to larger doping levels up to $E_{\mathrm{F}} = 29.3$~meV.
We plot the absorption spectra of the polaritonic system in Fig.~\ref{fig:spectra}, showing its modification at increasing $E_{\mathrm{F}}$. Here we assume mode broadening of $\gamma = 1$~meV being characteristic to modern setups.\cite{Emmanuele2020} Due to light-matter coupling with several modes simultaneously, we observe several distinct polaritonic modes with separate anticrossings. 
\begin{figure}
    \centering
    \includegraphics[width = 0.95\linewidth]{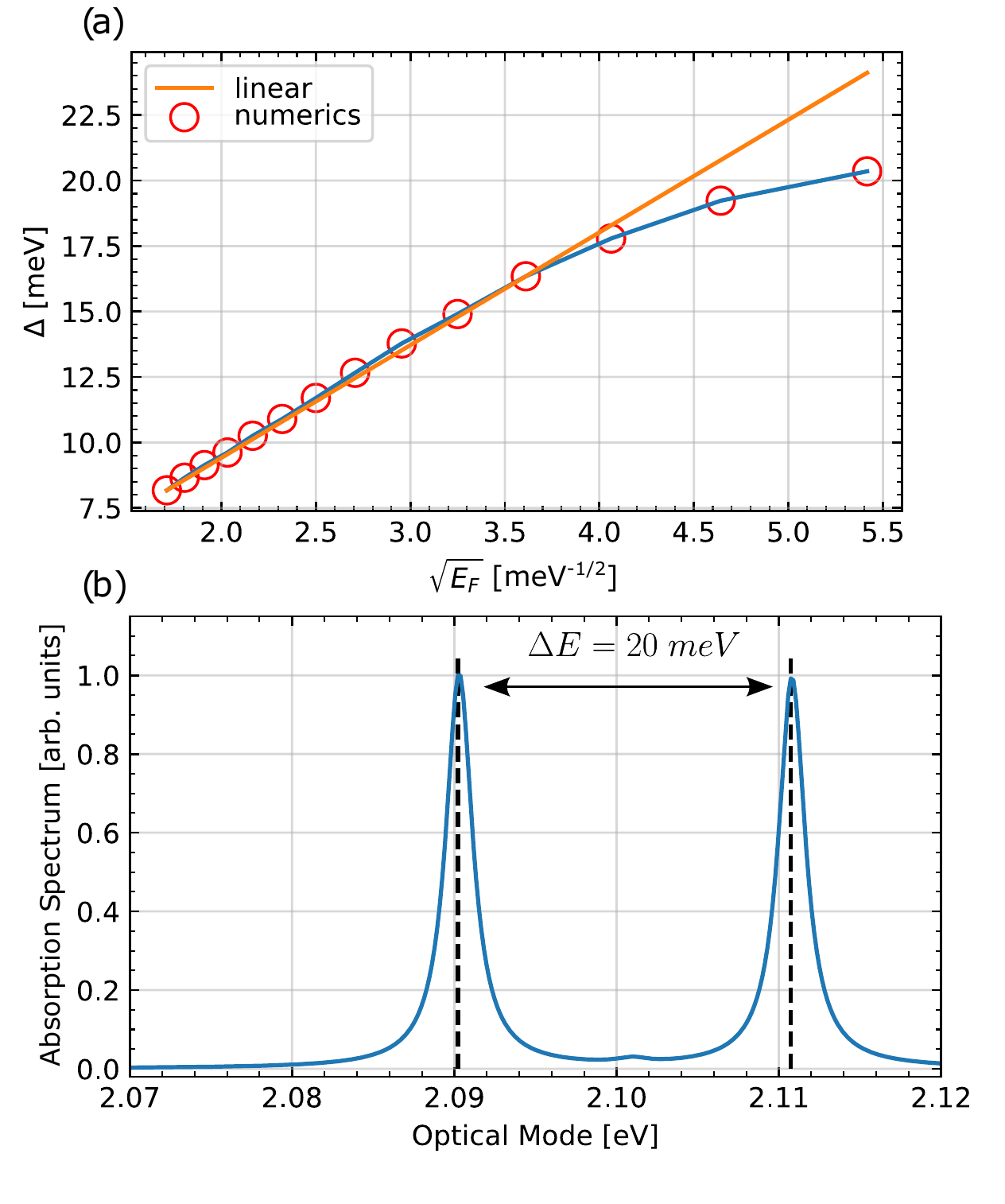}
    \caption{\textbf{Charge density distributions}. We show the distribution of various $\tau s_z=+1/2$ trionic states at $E_F=$10.56 meV: (a) -- trion state (attractive polaron), (b) -- indirect exciton state (c) -- direct exciton state (repulsive polaron) and (d) -- potential excited trion state. All four states are bright. Both excitons states exhibit Pauli blocking effect, which is observable at the top of the valence band.
    }
    \label{fig:wave}
\end{figure}
\begin{figure}
    \centering
    \includegraphics[width = 0.95\linewidth]{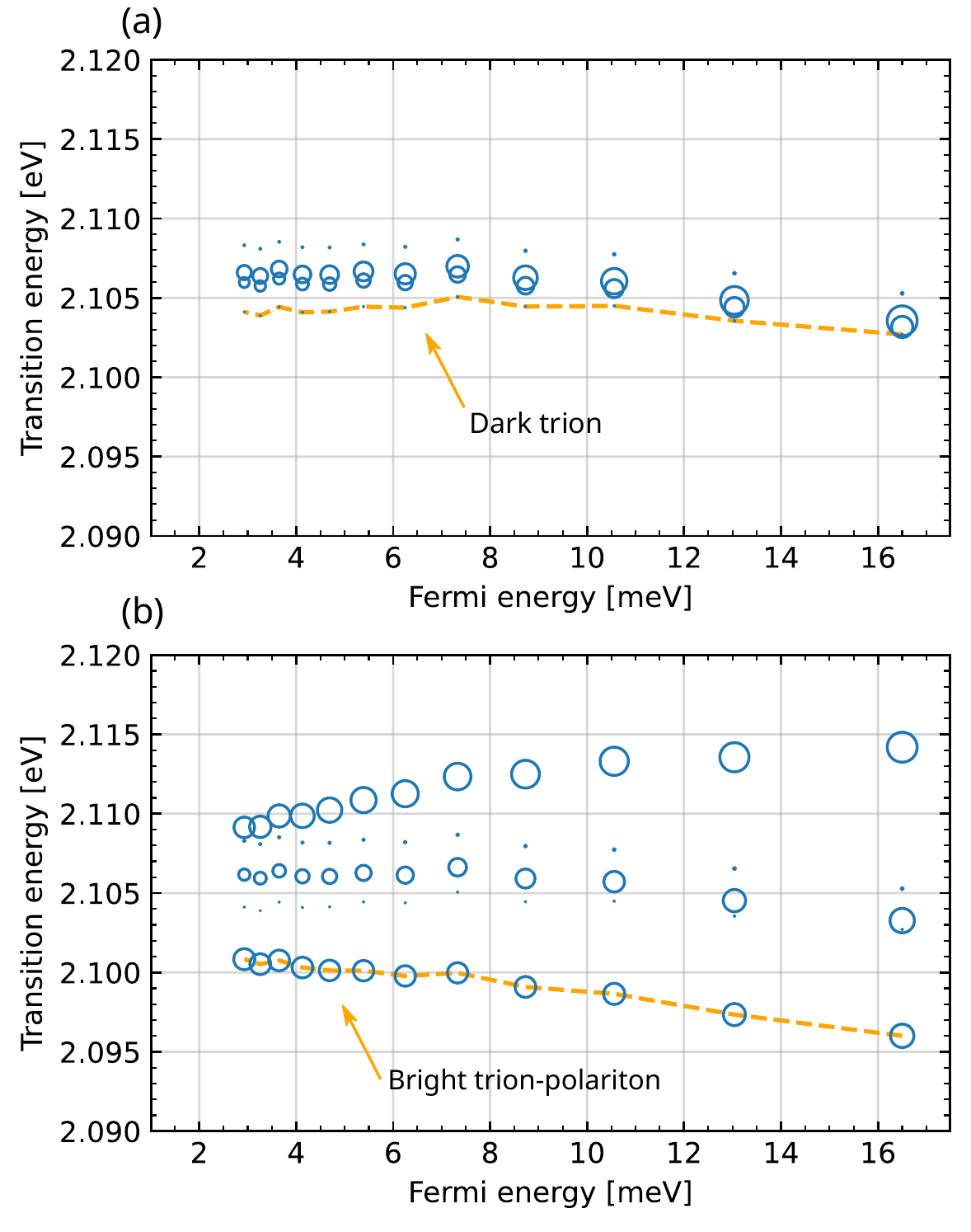}
    \caption{\textbf{Ground state brightening.} We observe the qualitative change of the ground state due to strong light-matter coupling. (a) Doping dependence of trion energies ($y$ axis) and their dipole moment (circle size). The lowest trion mode is dark (orange dashed line) for the uncoupled case, which leads to suppression of the photoluminescence. (b) Doping dependence of trion polaritons in MoS$_2$ monolayer, where cavity mode is tuned close in energy to trion modes. The cavity-modified ground state is bright, thus enabling efficient photoluminescence from the system. The cavity mode is at $E_0 = 2.14 \; eV$ and we consider linear polarization.}
    \label{fig:brightening}
\end{figure}

At low doping in [Fig.~\ref{fig:spectra}(a)], we observe two clear anticrossings between the cavity mode and three-body excitations, corresponding to bound trions and 1s exciton states renormalized by the interaction with free electrons.\cite{Rana2021,Li2021a} While the excitonic response dominates at higher energies, we note the significant oscillator strength for the low-energy trion modes and a clear sign of anti-crossing. Taking the lowest bright trion (attractive polaron) as an example, we check that the effective Rabi energy for this mode [Eq.~\ref{eq:Rabi_def}] is equal to $4.2$~meV at $E_{\mathrm{F}}=3.26$~meV, being larger than $1$~meV broadening. At the same time, we note that due to the multi-mode structure of our system the splitting between modes has to be estimated directly from the spectrum, as we show later in Fig.~\ref{fig:contributions}.

At intermediate doping in [Fig.~\ref{fig:spectra}(b,c)], the most pronounced effects are: (1) the increase of trion-polariton splitting; (2) qualitative change of exciton-polariton coupling. For point (1), we track the energy splitting at resonant detuning and plot it as a function of the square root of the Fermi energy (and electron concentration), $\sqrt{E_\mathrm{F}}$. There is a known linear scaling of the trion-polariton splitting~\cite{Rapaport2000,Rapaport2001,Ramon2003,Kyriienko2020}, and we confirm the expected scaling at small and intermediate doping levels [see details in Fig.~\ref{fig:contributions}(a)]. Specifically, we show how the trion energy splitting grows, comparing numerical results [red circles and blue curve in Fig.~\ref{fig:contributions}(a)] with the linear scaling [Fig.~\ref{fig:contributions}(a), yellow line]. For point (2), we note that additional mode that brightens with the increase of $E_\mathrm{F}$ was noted in Ref.~[\onlinecite{Zhumagulov2020}]. As it is located at the middle of the polariton gap, the anticrossing takes shape characteristic to dipolaritons, and features flat dependence on the cavity mode energy  [Fig.~\ref{fig:spectra}(c,d)]. This feature enriches the polaritonic spectrum, which was overlooked in the previous theoretical analysis~\cite{Tan2020,Rana2021}. 

As doping level grows to the point when Fermi energy becomes of the order of trion binding energy, i.e. $E_{\mathrm{F}} \sim 20$~meV, additional polaritonic modes emerge from excitonic modes dressed by the Fermi sea. We observe several modes that gain the oscillator strength at $E_{\mathrm{F}} = 10.56$~meV [in Fig.~\ref{fig:spectra}(e), see spectral features at around $2.18$~eV]. These are attributed to the excited trion states, as follows from the charge distribution analysis presented below and in Fig.~\ref{fig:wave}. Results obtained from our microscopic theory in the large doping limit have several remarkable features [Fig.~\ref{fig:spectra}(f)]. The trion splitting is large in absolute values and reaches $18$~meV value at $E_{\mathrm{F}} = 29.34$~meV [see spectrum cross-section in Fig.~\ref{fig:contributions}(b)]. At resonance with the trion mode, the cross-section shows two dominant peaks and weakly coupled excitonic modes at higher energies.  The splitting shows a sublinear dependence with $\sqrt{E_F}$, which can be attributed to the increase of the role of the effects of the Pauli blocking and redistribution of the oscillator strength. However, we stress that strong coupling to trions remains at all doping levels.

To understand the nature of quasiparticles, we plot the charge distribution for different states (Fig.~\ref{fig:wave}). This plot shows the contributions of single-particle states with different $k$ into the full three-particle wavefunction for four many-body states. All states in Fig.~\ref{fig:wave} are bright because all four quasiparticles have direct electron-hole pairs with equal spins.
The first state in Fig.~\ref{fig:wave}(a) represents the trion state, which can be seen from the symmetrical conduction bands occupation. The latter manifests the equivalence between two electrons in the three-particle state. 
The second state in Fig.~\ref{fig:wave}(b) represents the momentum-indirect exciton state. Indeed, we observe a fully localized exciton-state for electron and hole being in K$^\prime$ and K valleys, respectively, while the quasi-free electron is located in the K-valley (described by small charge density spreading in the reciprocal space). We also note a signature of Pauli-blocking in the valence band occupation at K point.
The third state, shown in Fig.~\ref{fig:wave}(c), is the direct exciton with the Pauli-blocking signature in the valence band occupation, plus quasi-free electron in the same valley. 
Finally, the fourth state in Fig.~\ref{fig:wave}(d) energetically is located above three previous states but does not exhibit Pauli blocking signature in the valence band occupation and has an $s$ orbital-like occupation. It features the charge distribution that corresponds to trion modes. We thus conclude that it is an optically active 2s trion state, which was experimentally observed~\cite{Arora2019}.  We have shown only a few states, which correspond only to the  $\tau s_z =+1/2$ states. However, there are much more optically active three-particle states.  For example, there are at least three optically active fully symmetrical trion states \cite{Zhumagulov2021}. We anticipate similar behavior for the indirect/direct excitons and the excited trion states.

An essential consequence of strong coupling in  MoS$_2$ monolayers is the qualitative change of the ground state properties. The three-body quasiparticles of a bare monolayer include four trionic modes, with the dark mode being energetically favorable [see Fig.~\ref{fig:brightening}(a)]. Qualitatively, this ground state corresponds to the suppressed photoluminesce from the uncoupled system. Similar physics is observed in carbon nanotubes, where tightly bound excitons have zero optical dipole moments \cite{Perebeinos2004}. The situation changes dramatically when the TMDC monolayer is placed in the optical cavity, where strong coupling modifies energies and redistributed dipole moments [Fig.~\ref{fig:brightening}(b)]. In this case, a pair of polariton modes formed from the bright trion, and the lower one can go beyond the energy of a dark trion uncoupled to light, as shown in Fig.~\ref{fig:sketch}. As a consequence, the ground state of the system now becomes optically active.


\section*{Conclusions}

In conclusion, we have developed a microscopic theory that describes strong light-matter coupling in doped TMDC monolayers in a wide range of free electron concentrations. The theory involves a numerical calculation of a set of three-body excitations, which are then coupled non-perturbatively to the cavity mode. We have calculated polariton spectra using this approach, which reveals the rich structure of the emerging composite light-matter quasiparticles. 
Our results confirm the robust nature of trion polaritons that holds in a large range of electron concentrations. At intermediate doping levels, we observe the three-mode structure of exciton-polaritons, where the presence of the additional mode in the middle of the polariton gap changes the anticrossing behavior qualitatively. Also, we predict that in polaritonic samples with large doping, the oscillator strength of the excited trion state becomes significant and may drive the system into a strong coupling regime with exotic charge distribution. Finally, we have demonstrated that increasing the electron concentration changes the nature of the system's ground state from dark to bright. Our results are crucial for building electrically tunable nonlinear polaritonic devices based on 2D materials.

\section*{Methods}

\small{
To calculate the energy spectrum and wavefunctions of many-body state we use the generalization of the Tamm-Dancoff approach \cite{Deilmann2016,Drppel2017,Torche2019,Tempelaar2019,Zhumagulov2020,Zhumagulov2020a}. We start by decomposing the full many-body Hamiltonian $\hat H$ in the three-particle state basis [Eq.~\eqref{eq:T_basis}]. Next, we solve the corresponding eigenvalue problem. 
In the trion basis, the full many-body Hamiltonian consists of several parts, $\hat H = \hat H_0 + \hat H_{c,c} + \hat H_{c,v}$, with $\hat H_0$ being the free-electron Hamiltonian, $\hat H_{c,c}$ the Coulomb coupling between electrons, and $\hat H_{c,v}$ the Coulomb coupling between electrons and holes. Explicitly, these terms read
\begin{align}
\label{eq:three_particle}
& {  H}  = {  H}_{0} + {  H}_{cc}  + {  H}_{cv},  \\
& {  H}_{0}= (\varepsilon_{c_1} + \varepsilon_{c_2} -\varepsilon_{v}) \delta_{c_1c_1^\prime} \delta_{c_2c_2^\prime}  \delta_{v v^\prime}, \nonumber \\
& {  H}_{cc} = (W_{c_1c_2}^{c_1'c_2'}-W_{c_1c_2}^{c_2'c_1'})\delta_{vv'}, \nonumber \\
& {  H}_{cv}= -(W_{v'c_1}^{vc_1'}-V_{v'c_1}^{c_1'v})\delta_{c_2c_2'}-(W_{v'c_2}^{vc_2'}-V_{v'c_2}^{c_2'v})\delta_{c_1c_1'} \nonumber \\
& \quad +(W_{v'c_1}^{vc_2'}-V_{v'c_1}^{c_2'v})\delta_{c_2c_1'}+(W_{v'c_2}^{vc_1'}-V_{v'c_2}^{c_1'v})\delta_{c_1c_2'}, \nonumber
\end{align}
where $W^{ab}_{a^\prime b^\prime} = W(\textbf{k}_{a}-\textbf{k}_{a^\prime})\langle a^\prime \vert a\rangle \langle b \vert b^{\prime} \rangle$ and $V^{ab}_{a^\prime b^\prime} = V(\textbf{k}_{a}-\textbf{k}_{a^\prime})\langle a^\prime \vert a\rangle \langle b \vert b^{\prime} \rangle$ are screened and bare Coulomb matrix elements, respectively. $\varepsilon_{c,v}$ are free particle energy terms for the conduction ($c$) and the valence ($v$) band.

Being eigenstates of the full many-body Hamiltonian, one can find the elements of $A$ by solving the eigenvalue problem
\begin{align}
\sum_{c_1^\prime c_2^\prime v^\prime} {  H}_{c_1c_2v}^{c_1^\prime c_2^\prime v^\prime} \bm{A}_{\nu, c_1^\prime c_2^\prime v^\prime} = \epsilon_\nu \bm{A}_{\nu, c_1c_2 v},
\label{eq:matrix-equation}
\end{align}
where matrix elements ${ H}_{c_1c_2v}^{c_1^\prime c_2^\prime v^\prime}$ are described in Ref.~[\onlinecite{Zhumagulov2020}.
A direct solution of Eq.~\eqref{eq:matrix-equation} is an extremely challenging task due to the high dimensionality of the trion Hamiltonian, which is proportional to $N_v N_c^2 N_k^2$, where $N_v$ is a number of valence bands, $N_c$ is a number of conduction bands, and $N_k$ is a number of k-points. For the converged calculations, the linear dimensions of the Hamiltonian matrix are more than $\sim 10^6$ states. In the case of  two-body exciton Hamiltonian, the dimensionality scales significantly slow as $N_v N_c N_k$. However, the trion Hamiltonian matrix is sparse, with about 1\% of the nonzero matrix elements. Therefore, we use the Arnoldi algorithm~\cite{Arnoldi1951}, which is implemented in ARPACK~\cite{ARPACK}, for obtaining low-energy eigenstates.
}

\section*{Data availability}

The data that support the findings of this study are available from the corresponding author upon reasonable request.

\section*{Code availability}
All the codes were implemented using home-built programs, which use standard open-source software packages.

\begin{acknowledgments}
 Y.\,V.\,Z. is grateful to the Deutsche Forschungsgemeinschaft (DFG, German Research Foundation) SPP 2244 (Project-ID 443416183) for the financial support. The work of D.\,R.\,G. related to the numerical calculation of the structure of three-body states was supported by Russian Science Foundation (project 20-12-00224). I.\,A.\,S. acknowledges support from Icelandic Research Fund (project ``Hybrid polaritonics'') and Russian Foundation for Basic Research (RFBR, joint RFBR-DFG project No. 21-52-12038). S.\,C. and O.\,K. acknowledge the support from UK EPSRC New Investigator Award under the Agreement No. EP/V00171X/1. V.\,P.  acknowledges support from the
Vice President for Research and Economic Development
(VPRED) at the University at Buffalo, SUNY Research
Seed Grant Program, and the Center for Computational
Research at the University at Buffalo~\cite{UBCCR}. 
\end{acknowledgments}

\section*{Author contributions} Y.\,V.\,Z. performed numerical calculations of trion spectra, with contributions from V.\,P. and D.\,R.\,G. S.\,C. performed calculations for polaritonic spectra and prepared the figures. O.\,K., I.\,A.\,S., and V.\,P. wrote the manuscript, with crucial contributions from all authors.

\section*{Competing interests} 
The authors declare no competing interests.


\end{document}